\documentclass[%
 reprint,
superscriptaddress,
 amsmath,amssymb,
 aps,
 prl,
]{revtex4-1}

\usepackage{soul}
\usepackage{graphicx} 
\usepackage{hyperref}
\usepackage{dcolumn}
\usepackage{bm}
\usepackage[mathlines]{lineno}

\usepackage{miller}
\usepackage{xcolor}
\usepackage[normalem]{ulem}
\begin{document}


\title{Parameter-free quantitative simulation of high dose microstructure and hydrogen retention in ion-irradiated tungsten}

\author{Daniel R. Mason}%
\affiliation{UK Atomic Energy Authority, Culham Science Centre, Oxfordshire OX14 3DB, UK}%
\email{Daniel.Mason@ukaea.uk}

\author{Fredric Granberg}
\affiliation{Department of Physics, University of Helsinki, P.O. Box 43, FI-00014, Helsinki, Finland}%

\author{Max Boleininger}
\affiliation{UK Atomic Energy Authority, Culham Science Centre, Oxfordshire OX14 3DB, UK}%

\author{Thomas Schwarz-Selinger}
\affiliation{Max-Planck-Institut f\"ur Plasmaphysik, Boltzmannstr. 2, 85748 Garching, Germany}%

\author{Kai Nordlund}
\affiliation{Department of Physics, University of Helsinki, P.O. Box 43, FI-00014, Helsinki, Finland}%

\author{Sergei L. Dudarev}
\affiliation{UK Atomic Energy Authority, Culham Science Centre, Oxfordshire OX14 3DB, UK}%

\date{March 2021}    
 
\begin{abstract}
Hydrogen isotopes are retained in materials for fusion power applications, changing both hydrogen embrittlement and tritium inventory as the microstructure undergoes irradiation damage.
But modelling of highly damaged materials - exposed to over 0.1 displacements per atom (dpa) - where asymptotic saturation is observed, for example tungsten facing the plasma in a fusion tokamak reactor, is difficult because a highly damaged microstructure cannot be treated as weakly interacting isolated defect traps.
In this paper we develop computational techniques to find the defect content in highly irradiated materials without adjustable parameters.
First we show how to generate converged high dose ($>1$ dpa) microstructures using a combination of the creation-relaxation algorithm and molecular dynamics simulations of collision cascades.
Then we make robust estimates of point defects and void regions with simple developments of the Wigner-Seitz decomposition of lattice sites.
We use our estimates of the void surface area to predict the deuterium retention capacity of tungsten as a function of dose.
This is then compared to $^3$He nuclear reaction analysis (NRA) measurements of tungsten samples self-irradiated at 290 K to different damage doses and exposed to deuterium plasma at low energy at 370 K. We show that our simulated microstructures give an excellent match to the experimental data, with both model and experiment showing  1.5-2.0 at.\% deuterium retained in tungsten in the limit of high dose.
\end{abstract}

\maketitle

\section{Introduction}

Materials intended for the first wall and divertor of the proposed DEMO fusion reactor are expected to face irradiation doses of 10~dpa or more depending on location~\cite{You2016,Gilbert_FST2014}, but in the absence of materials, which have actually experienced such high neutron fluxes with a fusion spectrum, it is expected that we will need to extrapolate device performance to some extent using modelling and simulation.
However, in recent years it has become increasingly clear that the methods developed to simulate microstructural evolution in the \emph{dilute} defect limit are not well suited to generate representative \emph{highly irradiated} microstructures. 
While it is correct to model an isolated prismatic dislocation loop as a diffusing single entity~\cite{Derlet_PRB2011,Arakawa_Science2007}, and substantial progress has been made with object kinetic Monte Carlo~\cite{MartinBragado_CPC2013,Domain_JNM2004,Stoller_JNM2008,Becquart_JNM2009,Castin_JNM2019} and cluster dynamics models~\cite{Marian_JNM2011,Liu_Acta2017} based on discrete defect objects with parameterized effective mobilities, this approach must break down in the limit where the defect concentration grows to the point when mean free paths are short or elastic interactions dominate the dynamics. 
Correlated defect motion is observed in simple theoretical models incorporating elastic interactions~\cite{Dudarev_PRB2010,Li_PRM2019}, which well describes the experimentally observed phenomena involving loop rafting~\cite{Yi_acta2016,ElAtwani_Acta2018} and self-pinning of defects~\cite{Mason_JPCM2014}, provided that defects are inhomogeneously distributed.
In the dense microstructure limit ($>0.1$~dpa) new phenomena appear - dislocation loops can merge together to the point where they are better described as regions of near-perfect crystal than as isolated platelets of interstitial point defects~\cite{Debelle_PRM2018,Mason_PRL2020}. This is a fundamental topological transition in irradiated materials between the point where dislocation objects can be treated as loops to the point where dislocations form a material-spanning network.
Small changes in stress due to the production or movement of defects can unlock barriers previously too high to be overcome, leading to an avalanche of defect coalescence or recombination over a wide spatial extent~\cite{Derlet_PRM2020}.
Thermal activation controls the long time evolution of complex microstructures~\cite{Ferroni_Acta2015}, but in the dense limit, their generation and description requires adequate handling of the large number of degrees of freedom involved in complex defect rearrangement~\cite{Gra20,Sand_JNM2018,Marinica_PRB2011,Mason_MSMSE2019}.

The dense microstructure limit does have one simplifying universal feature, namely that the saturation of physical phenomena has been observed. 
Materials irradiated to $>1$~dpa have shown saturation in thermal diffusivity~\cite{Reza_Acta2020}, lattice strain~\cite{Mason_PRL2020}, and hardness~\cite{Gaganidze_JNM2011}.
This suggests that additional cascade damage does not significantly evolve the defect distributions, and therefore we may have a chance to simulate this dynamic steady state even while modelling non-linear transient states remain difficult.
In this paper we exploit the experimental observations that the vacancy content of ion-irradated tungsten, as deduced using positron annihilation spectroscopy, shows saturation~\cite{Hollingsworth_NF2019}, as does its deuterium retention capacity \cite{Ogorodnikova_JNM2015,Wang_JNM2021} at low temperature and high dose ($>0.2$~dpa), and generate a steady state distribution of the vacancy content of highly irradiated tungsten with atomic resolution and no adjustable parameters.

In a future D-T fusion reactor, it will be necessary to keep a careful inventory of the tritium retained in the walls, both in terms of efficient fuel generation and minimising its retention at the point of decommissioning or a loss-of-coolant accident~\cite{Causey_JNM2002,Tynan_NME2017}.
Hydrogen mobility in tungsten is high, with an activation barrier of order 0.2~eV~\cite{Heinola_JAP2010,Holzner_PS2020}, and its enthalpy of solution is also high at order 1~eV~\cite{Frauenfelder_JVST1969,Heinola_PRB2010}. Any hydrogen isotopes in tungsten will therefore either quickly migrate to existing defects~\cite{Lu_NucFus2014}, influencing their evolution~\cite{Kato_NucFus2015,SchwarzSelinger_NME2018} or even self-inducing new defects~\cite{Liu_AIPA2013,Qin_JNM2015,Zayachuk_NucFus2013,Hodille_PhysRevMat2018}.
Of the possible trapping sites for hydrogen in tungsten, a surface - either exterior or interior - is generally considered the most binding \cite{Heinola_PRB2010}, with a binding energy of one hydrogen to the \hkl[100] surface 0.8-0.9~eV~\cite{Heinola_PRB2010}, and to a vacancy 1.4~eV~\cite{Kato_JPFRS2009,Heinola_PRB2010b}. Up to six hydrogen atoms can bind to a single vacancy at zero temperature.
By contrast, the binding of a hydrogen atom to an interstitial atom is 0.3~eV~\cite{Becquart_JNM2010}, and to an interstitial loop 0.7~eV~\cite{DeBacker_NucFus2017}, and so while hydrogen atoms can form a Cottrell atmosphere in the elastic fields around dislocations they are unlikely to be present in large concentrations at room temperature except under plasma loading conditions~\cite{DeBacker_PhysScr2017}.

Models for hydrogen retention typically start with the assumption of one or more defect trapping site types~\cite{Shimada_JNM2011,Ogorodnikova_JAP2015}, and consider the effective diffusion of hydrogen from the traps to the surface or into bulk~\cite{Kirchheim_SSP2004}.
Due to the difficulty of modifying these equations due to the nanoscale fluctuations in stresses observed in highly-damaged microstructures~\cite{Derlet_PRM2020}, here we opt to find a theoretical maximum retention level based on the surface area alone~\cite{Hayward_JNM2016}.

Previously two simulation methodologies have been employed to generate heavily irradiated microstructures at the atomic scale.
 
The first method is the CRA method, where instead of following full cascades, Frenkel pairs are directly inserted by removing randomly chosen atoms and replacing them in new, randomly chosen positions. The new configuration is then relaxed using conjugate gradients (CG) with appropriate elastic boundary conditions~\cite{Chartier_APL2016,Debelle_PRM2018,Derlet_PRM2020}. 
This is iterated many times to build up damage. As the canonical definition of displacements per atom (dpa) is the number of vacancies produced, we have a well-defined measure of the (canonical) dpa level after a number of algorithmic steps.
This method is very efficient to getting to high doses. However, there is no temperature and no cascade overlap effects, which have been seen to be important in tungsten~\cite{Byg19,Fel19}.
The number of defects obtained are therefore a theoretical maximum value, and are thought to overestimate observed experimental effects~\cite{Debelle_PRM2018,Mason_PRL2020}. 

The second method is to generate damage incrementally using a large number of overlapping molecular dynamics cascade simulations. One atom is given high kinetic energy, order kilovolts or more, and the system is evolved in time using molecular dynamics with thermostats and barostats appropriate for the boundary conditions.
This process is repeated as many times as can be afforded.
This has been done for several metals, and doses achieved are on the order of tenths of a dpa~\cite{Gra15,Byg18,Vel17,Gra20}. 
The results of these simulations agree with Rutherford Back Scattering Channeling measurements on similarly irradiated  samples~\cite{zha17}. However, as one needs to cumulatively add up the dose, over ten thousand consecutive cascade simulations would be needed to reach one dpa. This limits the dose range that can be investigated using this direct cascade simulation approach.

In order to investigate converged heavily irradiated microstructures, without the need of tens of thousands of direct cascade simulations, we combine these two methods. We carry out cascade simulations to transform the CRA pre-generated microstructures, in order to form atomic configurations with an effective dose up to the order of dpa. The convergence is demonstrated by extending expensive cascade-only simulations to high dose.
This convergence is found to be robust and significant, and it links together a purely static relaxation method for generating microstructure (CRA) with a dynamic method (MD), in a parameter-free way.

We emphasize that we do not claim the microstructures are thermally annealed. To our knowledge there does not exist any simulation technique yet for annealing a highly damaged system of thousands of vacancies and dozens of arbitrarily complex interstitial loops and dislocation lines in a million atom box to experimental time scales.

To compute the void content, we need new methods suited for highly-irradiated systems. 
We discuss how to generate optimal Wigner-Seitz (W-S) point defect counts using a strained and rotated reference crystal.
We show how to generate isosurfaces enclosing voids using this reference crystal. By this method we can separate vacancies in vacancy dislocation loops, a mismatch between the number of atoms and the number of reference lattice sites but with no empty space for a hydrogen atom, from vacancies isolated or in small clusters which act as strong trapping sites.

Finally we use the computed void surface area to estimate hydrogen retention in highly damaged structures, and discuss the relationship with experimental measurements.


\section{Methods}

\subsection{Creation-Relaxation Algorithm simulations}
\label{CRA}

We produce representative highly damaged microstructures for this study with doses up to 3~dpa using a two-step process - first using the Creation-Relaxation Algorithm (CRA)~\cite{Derlet_PRM2020} to generate high-dose, high-energy unrelaxed structures, then relaxing them using massively overlapping Molecular Dynamics (MD) cascade simulations.


For our purposes, we reused the CRA-generated atomistic structures produced for Ref.~\cite{Mason_PRL2020}. These were simulated using the MNB interatomic tungsten potential, known to predict good vacancy structures~\cite{Mason_JPCM2017}. The geometry of the box was 64x64 conventional cubic bcc unit cells in the $x$- and $y$- directions, and 200 unit cells in the $z$-direction. The strain was set to zero in $x$- and $y$- directions in the plane parallel to the surface of the sample, representing a constraint on the damaged layer due to the pristine substrate material below, and the homogeneous stress was kept at zero in the $z$-direction, reflecting the traction-free surface boundary conditions.
The simulations represent a section cut from a foil with a damaged top layer, such as might be produced by heavy ion bombardment. Maintaining zero stress in the $z$-direction represents the ability of the foil to expand outwards, without including explicit surfaces in the simulation.
The simulations were performed using LAMMPS~\cite{LAMMPS}.

As very large local stresses are generated by the build up of point defects, a single CG relaxation step can produce atomistic rearrangement over a large spatial extent~\cite{Derlet_PRM2020}. Dislocation loops are formed, as interstitials are inherently mobile under stress. At a dose of order 0.1~cdpa, the dislocation loops combine into a network, changing the apparent number of lattice planes in the system. Vacancies are relatively immobile in this simulation, hence voids do not form. 
Importantly, as the only relaxation is essentially downhill in energy, the CRA method does not easily overcome thermal barriers. The structures produced are typically too dense in defects~\cite{Mason_PRL2020}.

A computational approach similar to the massively overlapping cascade simulations used in other metals~\cite{Gra15,Byg18,Vel17,Gra20} was used in our investigation to achieve full microstructural convergence in the limit of high dose.
The simulations were carried out with the MD code PARCAS~\cite{gha97,nor97}, with an adaptive timestep to account for the high energy atom movements~\cite{nor95}. Electronic stopping was applied as a friction force on all the atoms with a kinetic energy over 10~eV~\cite{Sand_EPL2013}. 
The interatomic potential utilized was the same as that used in the CRA simulations~\cite{Mason_JPCM2017}.

A perfect bcc simulation cell was created with the size $64~\times~64~\times~200$ conventional cubic bcc unit cells, consisting of about 1,600,000~atoms. The cell was created with the correct lattice constant at 300~K. The box was then thermalized to 300~K with fixed box sizes in the two shorter directions and kept at zero stress in the longest dimension. After the initial relaxation and thermalization, consecutive impacts were initiated in the box as follows: (i)  A 10~keV cascade was initialized in the centre of the cell, with a Berendsen thermostat~\cite{Ber84} at the border atoms and no pressure control. This was simulated for 20~ps; (ii) The cell was then relaxed with thermostat on all atoms and a pressure control~\cite{Ber84} to keep zero pressure in the longest dimension, this simulation lasted for 10~ps; (iii) After the relaxation, the cell was shifted randomly over the periodic boundaries, in order for the next cascade to impact a different region and to obtain a homogeneous irradiation of the whole cell. (iv) This was repeated many times. Two independent runs were conducted to see possible stochastic differences; it was concluded that the differences were small. 

\begin{figure}
    \centering
    \includegraphics[width=0.7\linewidth,angle=-90]{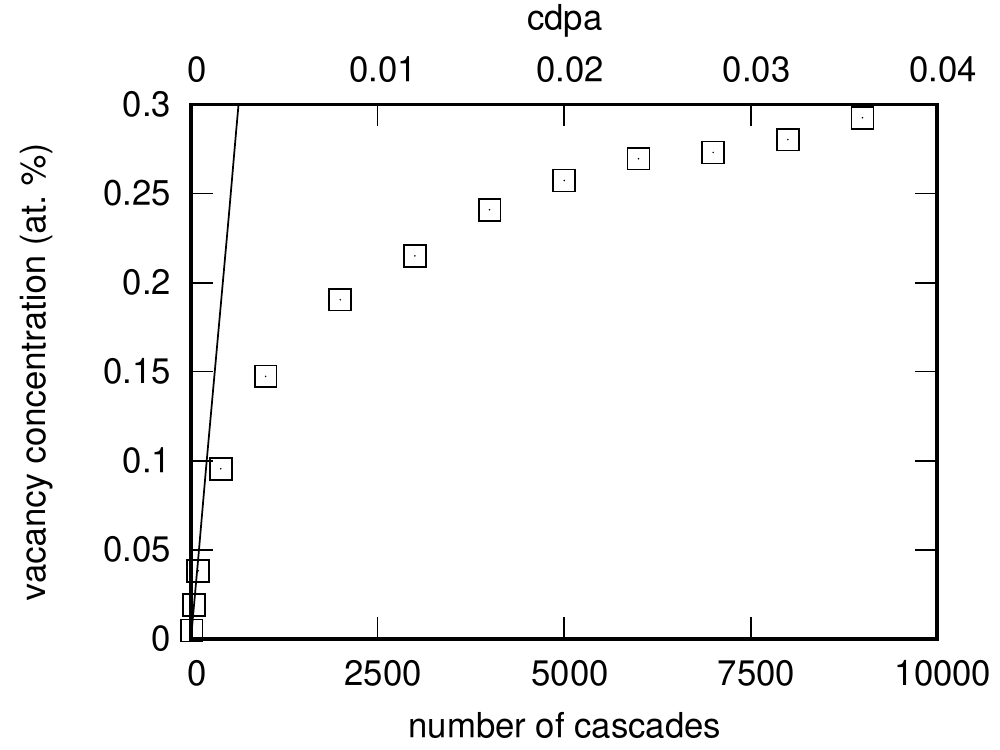}
    \caption{Vacancy concentration as a function of MD cascade count. The canonical dpa rate is the gradient of vacancy concentration with cascades at the origin, indicated by the solid line.}
    \label{fig:cascadeOnlyVsDPA}
\end{figure}

As the canonical definition of displacements per atom is the fraction of vacancies generated, the cdpa rate for the sequential cascades is given by the limit in increase in vacancy concentration, $c_v$, per cascade,
\begin{equation}
    \mathrm{cdpa} = \left. \frac{\partial c_v}{\partial N_{\mathrm{casc}}} \right|_{N_{\mathrm{casc}} = 0} \,  N_{\mathrm{casc}}
\end{equation}
In Figure~\ref{fig:cascadeOnlyVsDPA} we plot the vacancy content using the W-S method for our MD only cascade simulations. We estimate $cpda = 4.07\times 10^{-6} N_{\mathrm{casc}}$ using the first 40 cascades. 
We observe that the vacancy fraction reaches a concentration close to 0.3 at.\%. This is in line with previous experiments on tungsten~\cite{Reza_Acta2020} and other metals~\cite{Ave77}, as well as similar simulations on other metals~\cite{Gra15,Byg18,Gra20}. 
We extended these MD cascade simulations out to 10000 cascades, equivalent to 0.04~cdpa in a total simulation time of 300~ns.


The cascade annealing simulations followed the same procedure as described above.
The starting point were the highly damaged cells obtained via the CRA method~\cite{Derlet_PRM2020} and the cells generated in Ref~\cite{Mason_PRL2020}. These cells were initially scaled to the correct lattice constant (at 300 K) and thermalized to room temperature over 10 ps using a Berendsen thermostat~\cite{Ber84}, with the same boundary conditions as described above. After the initial thermalization, 1600 PKAs of 10 keV each were initiated in the cell. This corresponds to an additional dose of 0.0065 cdpa. 


\subsection{Void Detection}
\label{void_detection}

The Wigner-Seitz method is commonly used to identify the positions of point defects in crystalline materials in the low-damage limit~\cite{Nordlund_PRB1998}.
This technique imagines the Voronoi tesselation of an ideal lattice.
Atoms are placed into these Voronoi volumes, and the number of atoms in each Voronoi volume is counted separately.
If the atom positions are close to the reference lattice sites, each volume will be singly occupied.
If the atoms are significantly displaced from ideal lattice sites, then some volumes will have zero occupancy, others multiple occupancy. The volumes with zero occupancy are marked as vacancies.
In practice, the Voronoi geometry does not need to be calculated, it suffices to find the nearest reference site to each atom.

We are, in principle, free to choose our reference lattice. In the dilute defect case this is rarely a problem - we use the original perfect lattice far from the defect(s). In the dense defect case we need a well-defined method to find the reference.
Consider the simulation supercell to be a periodically repeating parallelepiped with repeat vectors $\vec{A}_1,\vec{A}_2,\vec{A}_3$.
By definition, for each point $\vec{x}$ there is an equivalent point $\vec{x} + N_1 \vec{A}_1 + N_2 \vec{A}_2 + N_3 \vec{A}_3$, where $N_{\alpha}$ are integers ($\alpha=\{1,2,3\}$).

Now consider a reference lattice of \emph{primitive} unit cells. This again is a periodically repeating parallelepiped with repeat vectors $\vec{b}_1,\vec{b}_2,\vec{b}_3$.
By definition, for each point $\vec{z}$ there is an equivalent point $\vec{z} + n_1 \vec{b}_1 + n_2 \vec{b}_2 + n_3 \vec{b}_3$, where $n_{\alpha}$ are integers.

For the two cells to be commensurate, it is necessary only for any triplet $\{N_1,N_2,N_3\}$ for there to exist a triplet $\{n_1,n_2,n_3\}$, such that 
    \begin{equation}
        N_1 \vec{A}_1 + N_2 \vec{A}_2 + N_3 \vec{A}_3 = n_1 \vec{b}_1 + n_2 \vec{b}_2 + n_3 \vec{b}_3.
    \end{equation}
Given the linearity of the problem, this reduces to 
    \begin{equation}
        \vec{A}_{\alpha} = n_{1\alpha} \vec{b}_1 + n_{2\alpha} \vec{b}_2 + n_{3\alpha} \vec{b}_3, 
    \end{equation}
or, in matrix notation $\mathbf{A} = \mathbf{b}\, \mathbf{n}$,
where $A_{\beta \alpha}$ is the $\beta^{th}$ Cartesian component of the $\alpha^{th}$ vector.
The matrix $\mathbf{n}$ is a matrix of nine independent integers, which has the flexibility to consider any combination of axial strains, shears and rotations necessary to fit the primitive unit cell into the simulation supercell.

Given the simulation cell $\mathbf{A}$, and an intention to use, say, a bcc primitive cell with $\vec{b}_1 = a_0 [\bar{1}11]$ etc, a fit for the number of unit cell repeats is
    \begin{equation}
        \label{eqn:bestFitRepeats}
        \mathbf{n} = \mathrm{nint}\left[ \mathbf{b}^{-1} \mathbf{A} \right],
    \end{equation}
where $\mathrm{nint}$ is the component-wise nearest-integer operator.
The best fit unit cell, strained and rotated appropriately, is then 
    \begin{equation}
        \label{eqn:bestFitVectors}
        \tilde{\mathbf{b}} = \mathbf{A} \mathbf{n}^{-1}.
    \end{equation}
Note that there is no requirement for $\mathbf{n}$ to remain constant through a simulation, particularly one with high-dose damage where new crystal planes could be formed. The counts of interstitials and vacant sites are not constrained to be equal.
    
We may have an estimate for the homogeneous rotation and strain in the system. This could be done by analysing the position of the peaks of the square of the structure factor $S(\vec{q}) = \sum_i \exp[ i \vec{q} \cdot \vec{x}_i ]$, where the sum runs over all atom positions $\vec{x}_i$.
Then we have a better starting estimate for the primitive cell $\mathbf{T}^0 \mathbf{b}$, and we can use this in place of $\mathbf{b}$ in equation~\ref{eqn:bestFitRepeats}.
We describe how we find a homogeneous transformation matrix $\mathbf{T}^0$ using a real-space method in the appendix.

The primitive unit cell has an associated motif $\vec{y}_i$, $i=\{1,2\ldots\}$ associated with it. For the bcc and fcc cases there is only one motif point, but hcp and diamond structure have two. The lattice is invariant under translation of the motif. In the dense defect case, where displacements may be large, there can arise a disregistry between reference lattice and displaced planes of atoms. This will typically manifest as smooth planar regions of point defects.
We can therefore say that the optimal translation for the Wigner-Seitz reference is the one with \emph{fewest point defects}.
This is a global minimisation problem, hard to solve generally, as an arbitrarily small displacement may push an atom from matching one reference site to another.
We find a good lattice offset with a simple grid search; we search $6 \times 6 \times 6$ trial offsets on a uniform grid, establish the one with the fewest point defects, then refine the grid and search again. This process is repeated a third time, giving a reasonably high-precision estimate for a minimum.

Finally we note that this procedure will still fail if there are multiple grains or phases in the system.
We used common neighbour analysis~\cite{Honeycutt_JPC1987,Stukowski_MSMSE2012} to search for subgrains in our simulation boxes, but did not find any.
Therefore we leave multi-grain defect detection for future research.

\begin{figure}
    \centering
    a)
    \includegraphics[width=0.75\linewidth] {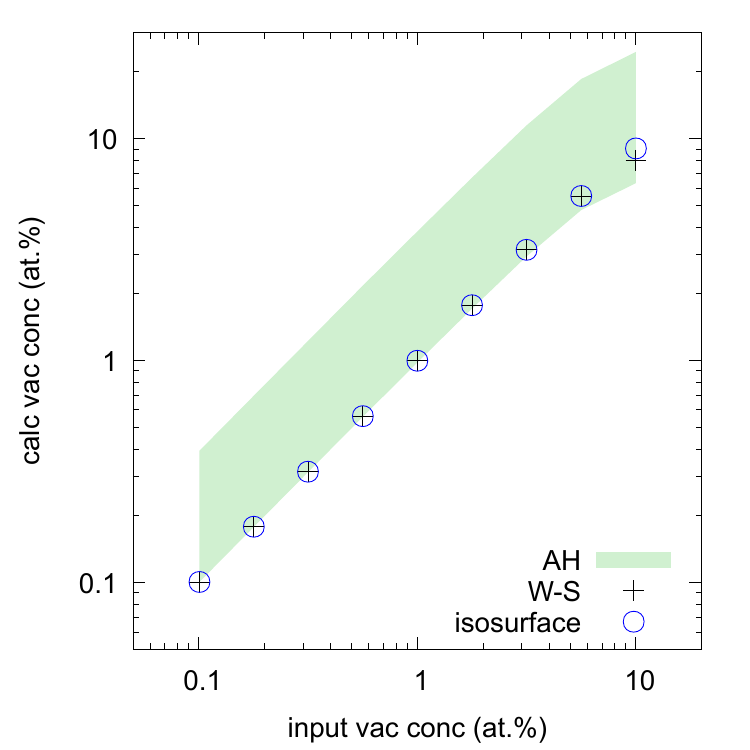}\\
    b)
    \includegraphics[width=0.75\linewidth] {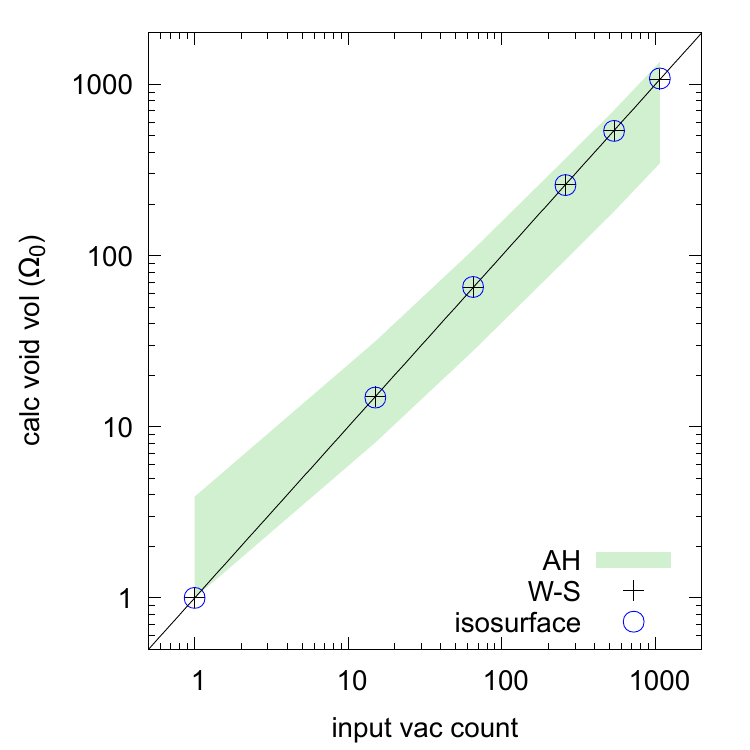}
    \caption{
    A comparison of three void detection methods - alpha hull (AH), Wigner-Seitz (W-S) and the isosurface method developed here.
    a) Performance of void detection algorithms with homogeneously distributed  vacancies. b) Performance with single spherical void. Solid line indicates perfect vacancy counting. Note the W-S algorithm performs perfectly in these cases.  }
    \label{fig:performance1}
\end{figure}

The successful operation of the (WS) method to determine vacancy point defects is illustrated in figure~\ref{fig:performance1}.


An alternate choice for void detection is to construct the alpha-hull (AH) from the Delauney tetrahedralisation of the atomic positions.

\begin{figure}
    \centering
    \includegraphics[width=\linewidth] {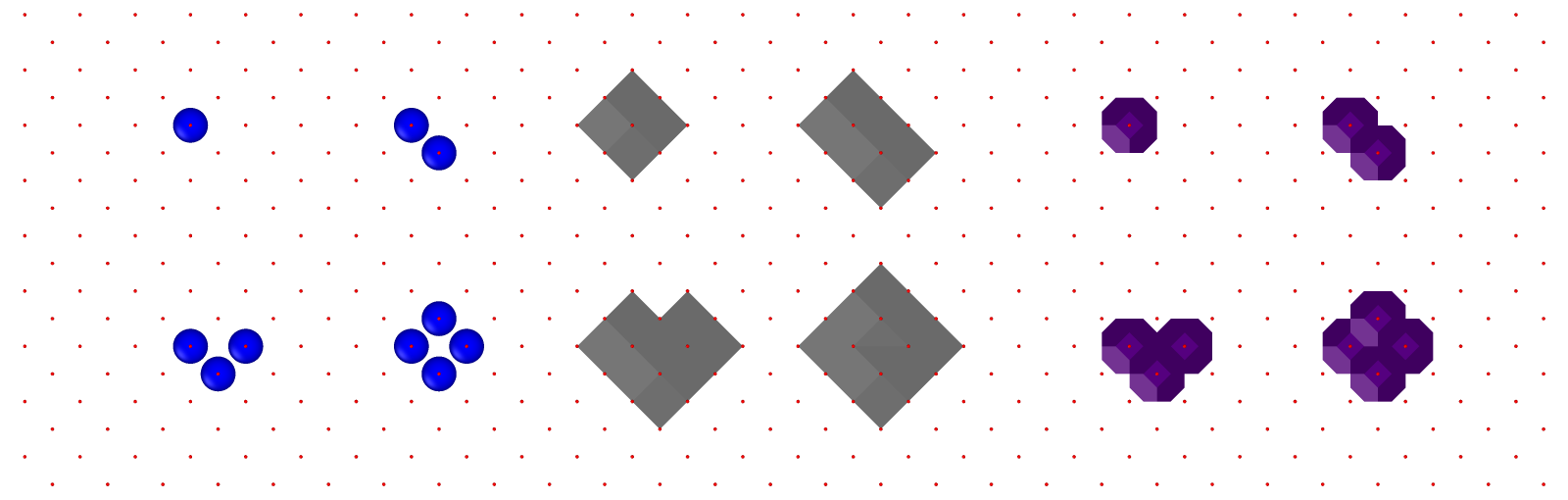}
    \caption{Small vacancy clusters identified using Wigner-Seitz occupation (left), Delauney triangulation (AH method) (centre) and Wigner-Seitz isosurfaces, described here (right)}
    \label{fig:comparison}
\end{figure}

This method requires a single input parameter, a minimum sphere radius used to determine concave surfaces, and so is highly transferable across atomic structures.
It is a robust method for determining the number and location of voids within a body, and is implemented in Ovito as the Construct Surface Mesh modifier~\cite{Stukowski_MSMSE2009}.
It does, however, produce surfaces with nodes defined by atomic positions. The volume of a monovacancy region is therefore overestimated, and the surface produced does not have the symmetry of the Wigner-Seitz cell.
The void-bounding surfaces produced for small vacancy clusters are illustrated in figure~\ref{fig:comparison}. Note that for this bcc example, the void surface for a monovacancy is dodecahedral, and encloses a volume $4 \Omega_0$.
Volumes and surface areas of larger vacancy cluster regions can also be reported, but with no simple linear transformation to determine the number of point defects represented.
In figure~\ref{fig:performance1} we illustrate this issue by representing the vacancy count using the AH method as a band. The lower limit assumes one vacancy has volume $4 \Omega_0$, and is correct for homogeneous distributions. The upper limit assumes one vacancy has volume $\Omega_0$, and is asymptotically correct for large voids.
 
Here we develop a new method which sits between W-S and AH.
If we know the lattice type, we can define vectors to the \emph{expected} positions of neighbours.
Writing $\vec{x}_i$ as the position of atom $i$, and $\vec{v}_{k,i}$ as the vector from $i$ to the expected location of neighbour $k$, the $k^{\mathrm{th}}$ Wigner-Seitz plane passes through the midpoint of $\vec{x}_i$ and $\vec{x}_i+\vec{v}_{k,i}$, and so is defined by the plane
    \begin{equation}
        \label{eqn:WignerSeitzNormal}
            \left( \frac{2\, \vec{v}_{k,i} }{ |\vec{v}_{k,i}|^2 } \right) \cdot \left( \vec{x} - \vec{x}_i \right) = 1.
    \end{equation}
Defining the $k^{\mathrm{th}}$ normal vector as $\vec{n}_{k,i} = 2\, \vec{v}_{k,i} / |\vec{v}_{k,i}|^2  $, we can find a scalar field $\psi_i(\vec{x})$ defining the distance from atom $i$ preserving the symmetry of the Wigner-Seitz cell:
    \begin{equation}
        \label{eqn:WignerSeitzVolume}
        \psi_i(\vec{x}) = \mathrm{max}_k \left\{ | \vec{n}_{k,i} \cdot \left( \vec{x} - \vec{x}_i \right) | \right\}.
    \end{equation}
The region closest to atom $i$ in the perfect lattice satisfies $\psi_i(\vec{x}) < 1$.
We can then define a scalar field $\phi(\vec{x})$ describing distance from any atomic position as the minimum value of $\psi_i(\vec{x})$ 
    \begin{equation}
        \label{eqn:phaseField}
        \phi(\vec{x}) = \mathrm{min}_i \left\{ \psi_i(\vec{x}) \right\}.
    \end{equation}
In an ideal lattice, this will have the value zero at atomic positions, rising to one at the Wigner-Seitz cell boundary.
If there is a void, then $\phi(\vec{x})>1$ in this region.
We illustrate a $\phi=1$ isosurface bounding small vacancy clusters in figure~\ref{fig:comparison}. 
For bcc we see the isosurface is a truncated octahedron, the correct shape for the Wigner-Seitz cell.

If the lattice is not ideal, but instead has a local strain tensor and rotation, then the vectors to neighbours can be found from those in the reference lattice, $\{\vec{v}^{0}_{k,i}\}$, by
    \begin{equation}
        \label{eqn:rotostrain_normal}
        \vec{v}_{k,i} = \mathbf{T} ( \vec{x}_i ) \vec{v}^{0}_{k,i} ,
    \end{equation}
where $\mathbf{T} = \left( \mathbf{I} + \mathbf{\epsilon} \right) \mathbf{R}$ is a combined rotation and strain at point $\vec{x}_i$.
We describe how to compute a converged local strain field $\mathbf{T} ( \vec{x} )$ in the appendix.

To compute the isosurface at $\phi=1$ in practice, we note that thermal fluctuations might make small `cracks' appear at isosurface level $\phi=1$, simply because atoms are instantaneously further away from each other than expected.
To compensate for this, we can compute the volume and area at isosurface level $\phi = 1 + \varepsilon$, where $\varepsilon$ is a small parameter of the order of the (fractional) vibration lengthscale of the atoms. This effectively redraws the Wigner-Seitz planes slightly further distant from the atoms, shrinking void regions proportionately. We therefore also compute the volume and area derivatives of the isosurface with respect to $\varepsilon$, and extrapolate to find the volume and area at exactly $\phi=1$. Volume and area are not sensitive to the choice of $\varepsilon$, we find taking $\varepsilon = 0.05$ works well in all cases we have tested.
The correct performance of our isosurface algorithm for simple homogeneous vacancies and voids is demonstrated in figure~\ref{fig:performance1}, where we show the calculated vacancy concentration for given input vacancy concentrations for W-S, AH and our isosurface method.

\begin{figure}
    \centering
    a)
    \includegraphics[width=0.75\linewidth] {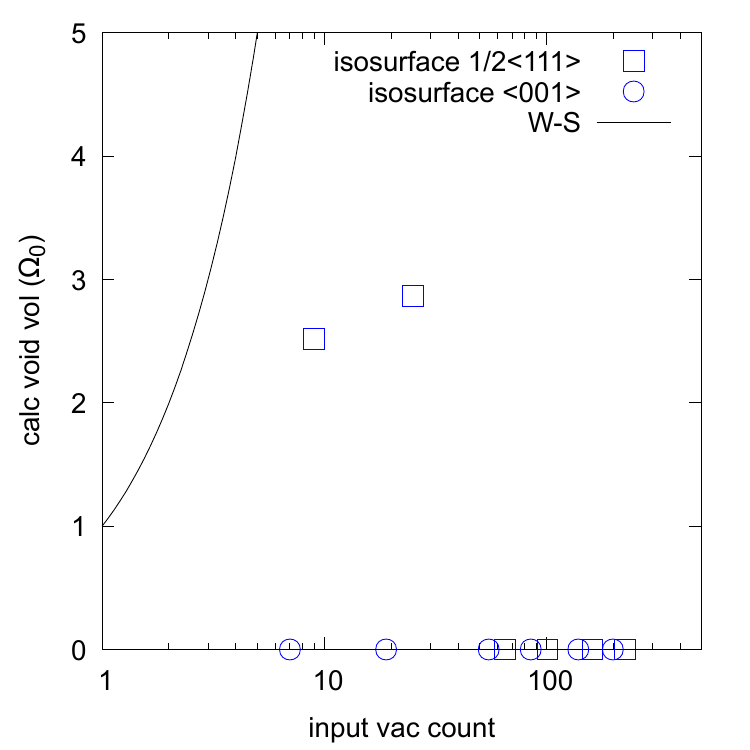}\\
    b)
    \includegraphics[width=0.6\linewidth] {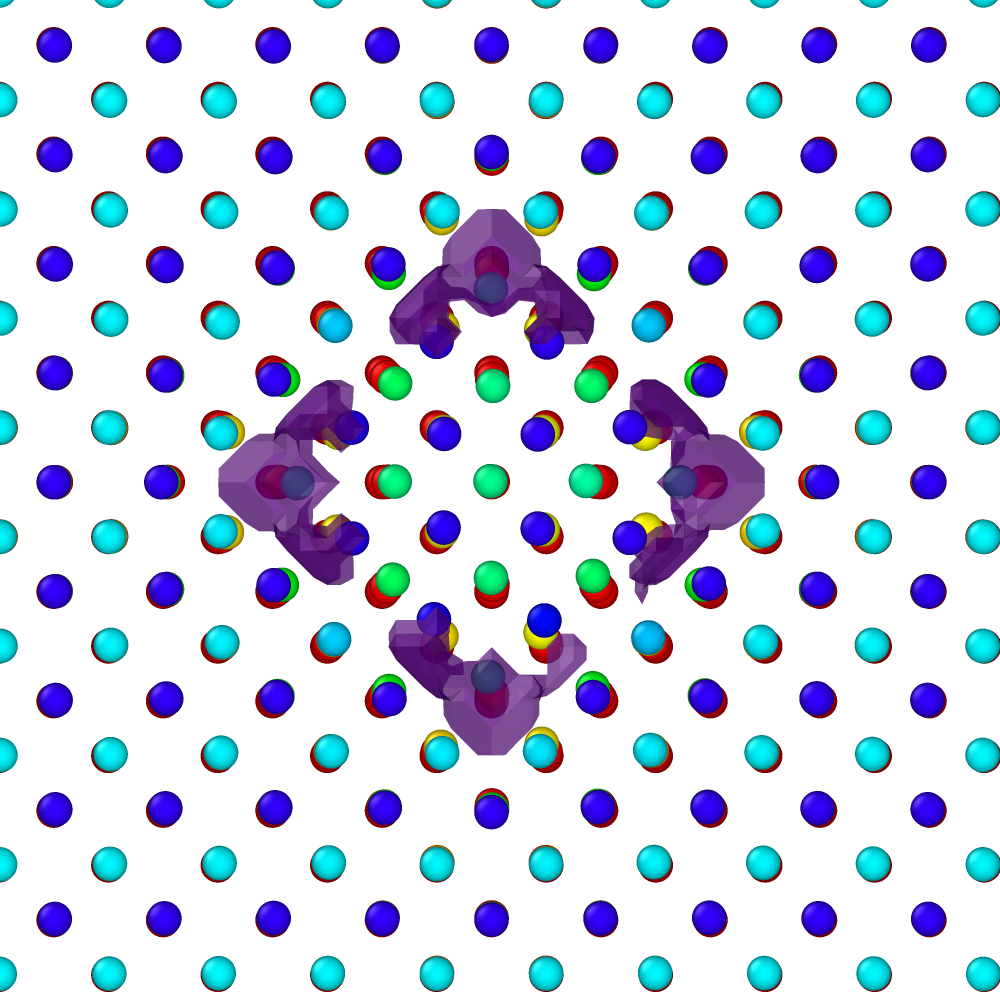}
    \caption{A comparison of the Wigner-Seitz (W-S) algorithm to the isosurface method developed here when applied to vacancy loops.
    a) Performance of void detection algorithm with prismatic vacancy loops in bcc tungsten. Two burgers vectors ($\mathbf{b}=1/2\langle111\rangle$ and $\mathbf{b}=\langle001\rangle$) are considered. Note that the W-S algorithm returns the number of vacant lattice sites exactly, whereas the isosurface method usually returns zero void space. b) slice through a $[001]$ vacancy loop viewed in the [001] orientation containing 25 vacant lattice sites, atoms coloured by depth. Isosurface showing the position of four voids overlaid.}
    \label{fig:performance2}
\end{figure}

In figure~\ref{fig:performance2} we demonstrate where our algorithm differs from W-S. For prismatic vacancy loops with Burgers vectors $\vec{b}=1/2\langle 111\rangle$, we find no void space, whereas W-S finds a difference between number of lattice points and atoms. For prismatic vacancy loops with $\vec{b}=\langle 001\rangle$ relaxed using the MNB potential~\cite{Mason_JPCM2017}, we find small voids opening at the dislocation core. 

\begin{figure}
    \centering
    \includegraphics[width=0.9 \linewidth] {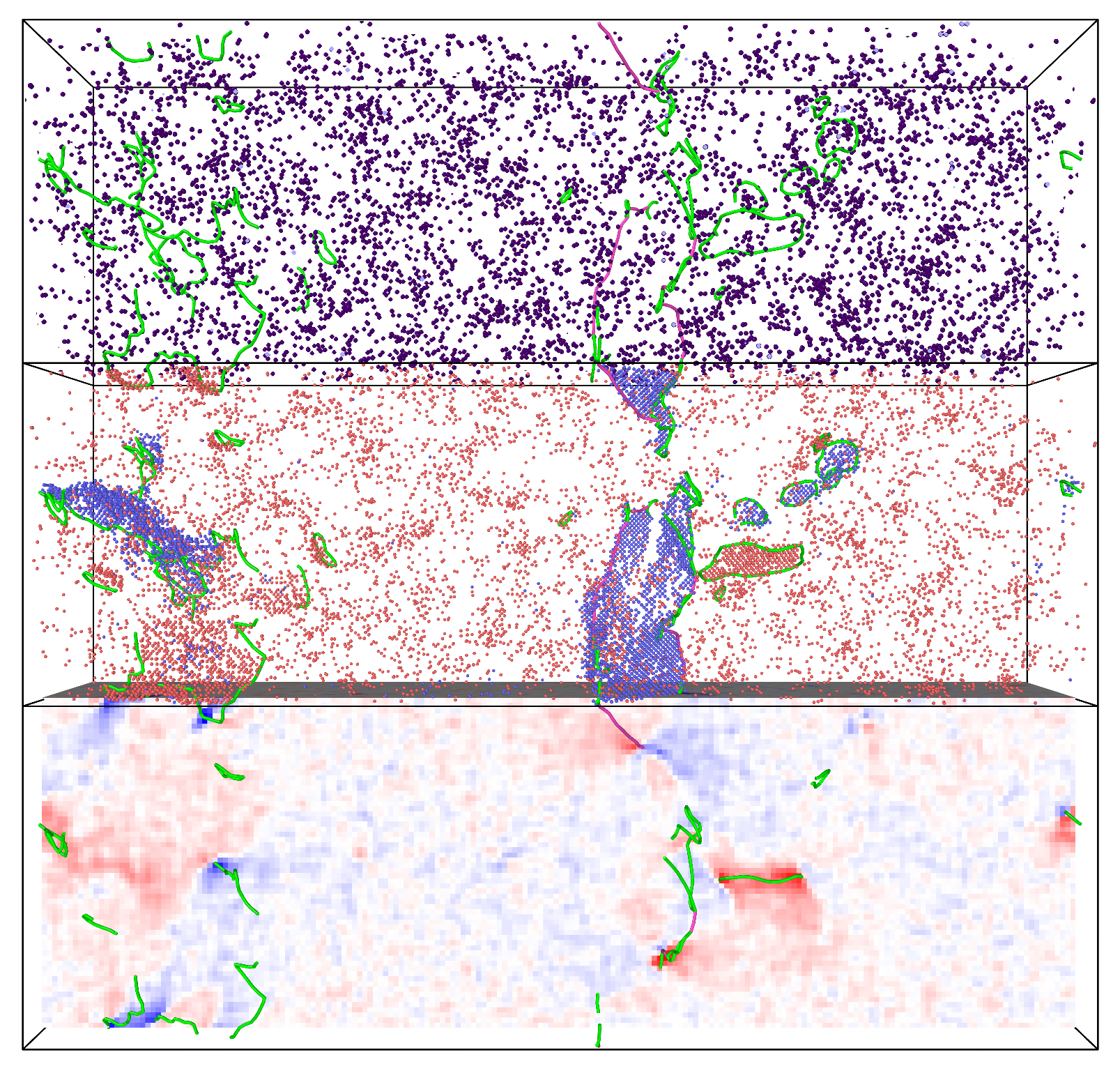}
    \caption{A simulation cell generated with CRA simulations to 1 dpa, then relaxed with MD cascades. The periodic replicas show (top) Void detection using isosurfaces, (centre) Wigner-Seitz point defect detection shows vacancies(red) and interstitials(blue), (bottom) Local strain calculation showing the $\epsilon_{xz}$ component coloured from blue (compressive -5\%) through white (no strain) to red (tensile +5\%). A dxa analysis\cite{Stukowski_MSMSE2012} is overlaid to show the dislocation loops and network. Green lines have burgers vector $1/2\langle 111 \rangle$ and pink lines $\langle 100 \rangle$. 
    Visualisation performed with Ovito \cite{Stukowski_MSMSE2009}.
    }
    \label{fig:high_dose_simulation}
\end{figure}

Figure~\ref{fig:high_dose_simulation} illustrates one of our CRA + MD simulation cells at 1~cdpa, using the defect detection algorithms described here. We can see a one-to-one match between the voids detected using our isosurfaces method and monovacancies/small vacancy clusters detected with the Wigner-Seitz method, except for at the vacancy dislocation loops. These are readily identified as planar features of higher `point defect' density in the central cell circled by dislocation lines. Note that some isosurfaces cross the periodic boundaries and appear as flecks. The bottom replica shows a slice through the local strain calculation, generated to improve the local estimation of the Wigner-Seitz normal vectors. 

\subsection{Experimental methods}

Tungsten samples with nominal purity 99.97~at.\% were procured from Plansee~\cite{Plansee}, outgassed, and annealed at 2000~K for 3 minutes, producing large grains 10-50~$\mu$m diameter and a low dislocation density estimated at $2\times10^{10}$~m$^{-2}$~\cite{Manhard_PractMet2015}.
The samples were irradiated to different damage dose ranging from 0.001 to 2.3~dpa at the peak maximum using 20.3~MeV W$^{6+}$ self ions at room temperature at the TOF beamline of the 3~MV tandetron accelerator at the Max-Planck-Institut für Plasmaphysik. Details can be found in Ref~\cite{Schwarz-Selinger_NME2017}. Damage doses were estimated using the Fast Kinchin-Pease option using SRIM~\cite{SRIM,Stoller_NIMB2013}, with a threshold displacement energy of 90~eV.

The samples were then exposed simultaneously to a low-temperature deuterium (D) plasma to decorate the produced defects with a moderate ion flux of $5.6 \times 10^{19}$~D~m$^{-2}$s$^{-1}$. For the chosen plasma parameters this ion flux hitting the samples contains dominantly D$_{3}^{+}$ ions with an energy below 15~eV. The target holder temperature was set with a liquid-cooled thermostat to 370~K, which is adequate to allow D diffusion into the bulk, and hence into the traps, but does not cause evolution of the traps themselves~\cite{Kapser_NF2018}.
For these exposure conditions the D retention is trap-limited, rather than diffusion limited, and the solute D atoms are not in equilibrium with the trapped population.
A full description of this implantation methodology can be found in Refs~\cite{Manhard_PlasmaSourSciTech2011,Schwarz-Selinger_NME2017}.

The retained deuterium population was analysed \emph{ex situ} using the D($^3$He,p)$\alpha$ nuclear reaction, with eight different $^3$He energies ranging from 500 to 4500~keV chosen to probe the sample depth up to 7.4~$\mu$m. 
A full description of the NRA analysis methodology used is given in Ref~\cite{SchwarzSelinger_NME2018}.


\section{Results and Discussion}
\label{results}

\begin{figure}
    \centering
    \includegraphics[width=0.7\linewidth,angle=-90]{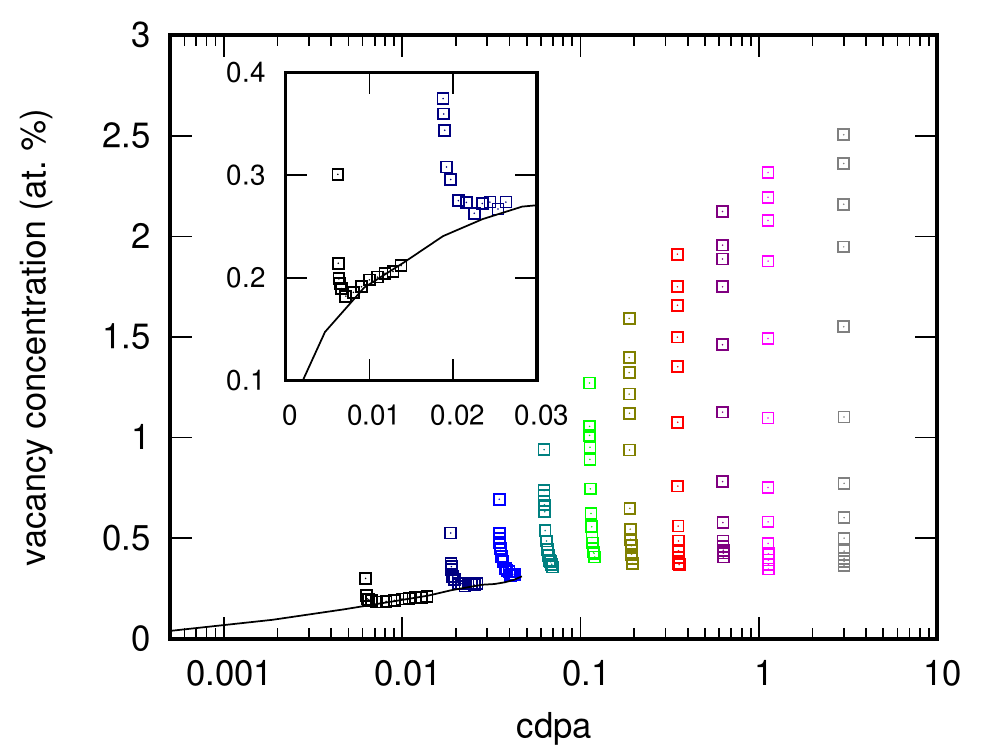}
    \caption{CRA simulations annealed with MD cascades. Open symbols: CRA simulations with MD annealing. Each coloured set starts with zero MD cascades at the highest vacancy fraction, and gradually decreases as MD cascades are added. Solid line: MD cascades only. Inset: same data plotted on a linear scale shows a match between cascade only and CRA+MD simulation techniques.}
    \label{fig:CRAplusCascadeVsDPA}
\end{figure}

The results of the combined CRA+MD cascade method is shown in Fig.~\ref{fig:CRAplusCascadeVsDPA}. In the figure, the solid line is the massively overlapping cascade simulations and the squares the CRA+Cascade annealing results. The CRA+Cascade annealing results are shown at these increments: before any cascade and after 10, 25, 50, 100, 200, 400, 600, 800, 1000, 1200, 1400 and 1600 cascades. Note that these increments are not linear, and Fig.~\ref{fig:CRAplusCascadeVsDPA} shows a dramatic change in vacancy concentration in the first few cascades, which ultimately slows to convergence.

Firstly, looking at the three lowest starting points, lowest doses in the CRA method, we can clearly see the cascade annealing effect, reducing the fraction of vacancies. This reduction continues until the solid line (MD-only simulations) is reached. We observe that as the number of cascades increases past this point, so does the vacancy fraction, in excellent agreement with the solid line. This validates that the combined method is resulting in the same defect concentration as the massively overlapping cascade simulations, however, here already with a speed up of a factor 2-5 depending on the CRA starting point. 

Secondly, looking at the high dose starting points, we again can observe a huge effect of cascade annealing in the beginning, with a small effect at the end of the run. We see a saturation of vacancy concentration at a level of 0.3~at.\%. With the direct validation against massively overlapping cascades at the lower doses and with the qualitative agreement (both defect concentration level and overall behaviour), we have shown this combined technique to be very effective on obtaining microstructures at very high doses - the microstructures  obtained by combining CRA + MD can result in a dose which would require about 1 million overlapping cascade simulations. This represents a dramatic speed-up on the order of 5000.

\begin{figure}
    \centering
    \includegraphics[angle=-90,width=0.9\linewidth] {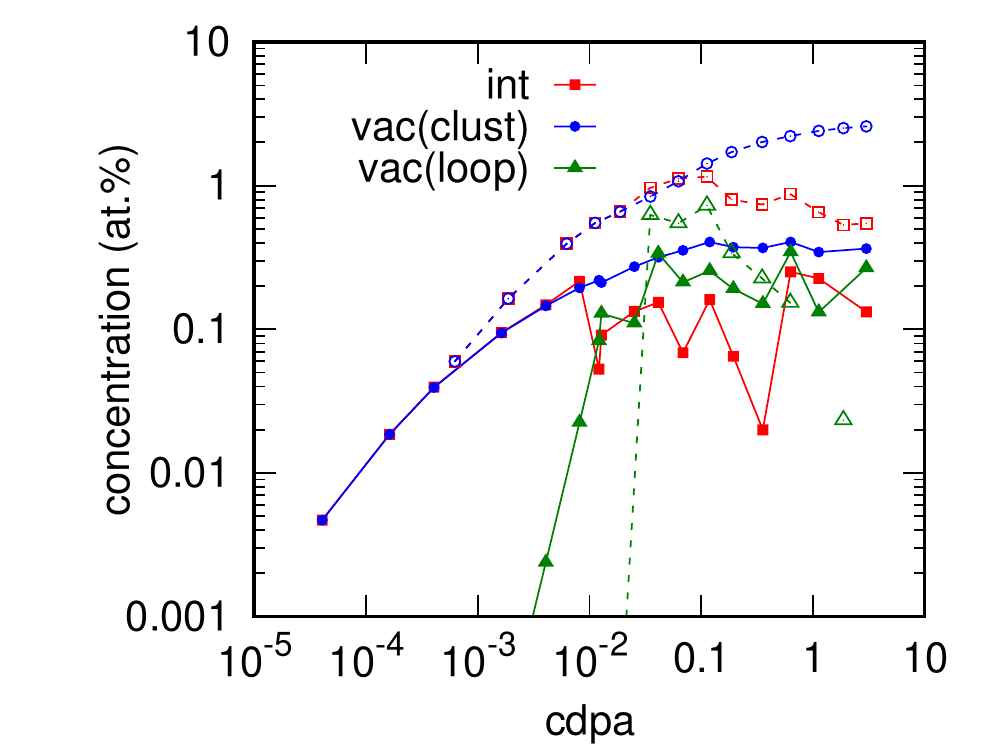}\\
    \caption{
    A count of point defects for high dose microstructure simulations.
    Interstitials counted using W-S method. Vacancies in clusters counted using isosurface method, and vacancies in loops taken as difference between total vacant lattice sites (W-S) and vacancies in clusters.
    Open symbols- result using CRA simulations alone.
    Solid symbols- result using CRA+MD relaxed simulations.
    The lines are to guide the eye between sets of points.
    }
    \label{fig:performance3}
\end{figure}

The resulting counts of all point defect types determined for our high-dose simulations is shown in figure~\ref{fig:performance3}. 
We find the interstitial count, and the total count of vacant lattice sites using the W-S method.
The proportion of vacancies in vacancy loops is found by subtracting the vacancies counted using the isosurface method from the total Wigner-Seitz count of vacant lattice sites.
We see that the atomic fraction of vacancies increases linearly with dose at low fluence, but tends to saturate at high fluence, with CRA at 3~at.\% and CRA+MD at 0.3~at.\%. 
This order of magnitude difference is clearly very significant for predicting properties of highly irradiated materials, and demonstrates the importance of relaxing CRA simulations.
The interstitial count is seen to follow the vacancy count at low fluence - this is an expected consequence of defects being generated as Frenkel pairs. But at high fluence interstitial dislocation loops extend across the periodic boundaries of the cell and form complete planes of atoms, with residual network dislocations.
For the simulation in Figure~\ref{fig:high_dose_simulation}, the matrix of primitive cell repeats is 
\begin{equation}
    \mathbf{n} = \left( \begin{array}{ccc}
                        0   &   64  &   201     \\
                        64  &   0   &   201     \\
                        64  &   64  &   1
                    \end{array} \right),
\end{equation}
indicating that, compared to the original zero dose simulation box, an additional plane normal to the $z$-direction with Burgers vector $1/2 \langle 111 \rangle$ has been added.

It is notable that this split between interstitial and vacancy count occurs earlier in the relaxed simulations, at 0.01~dpa compared to 0.1~dpa in the CRA only simulations. This occurs because the MD simulations provide sufficient energy for defect clusters to overcome thermal barriers.
We also see that vacancy loops emerge naturally in both CRA and CRA+MD simulations, and with a similar atomic fraction, at the point where interstitial and vacancy counts diverge. They can be clearly seen in figure~\ref{fig:high_dose_simulation}. These vacancy loops are generated by \emph{interstitial} loop coalescence: when the interstitials coalesce to form a plane they will not do so with 100\% coverage, but rather will leave small gaps. These gaps remain bounded by edge dislocation lines. They are, by definition therefore, vacancy loops.

%

\begin{figure}
    \centering
    \includegraphics[width=0.7\linewidth,angle=-90]{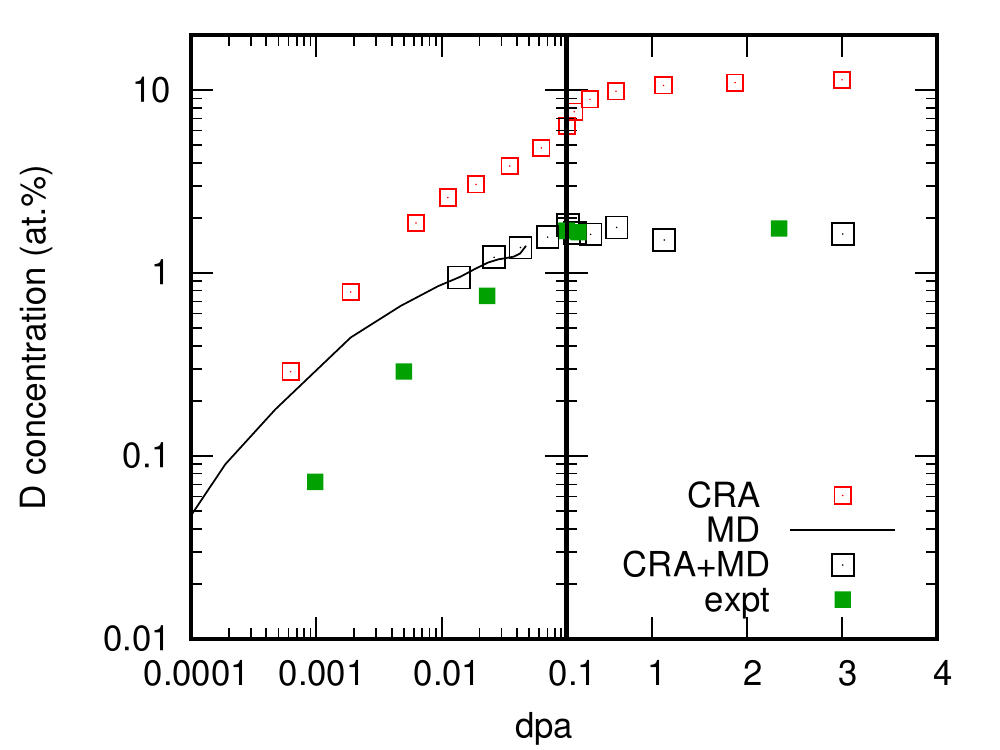}
    \caption{Deuterium (D) concentration assuming 5 D atoms per monovacancy equivalent. Note the displacement damage scale $x$-axis is split into logarithmic and linear halves, in order to emphasize the saturation level in the high dose limit.}
    \label{fig:Concentration}
\end{figure}

In order to make an estimate for the amount of hydrogen that can be retained in the material from the defect concentrations computed above, we need to estimate the trapping efficiency of the defects in our simulated microstructure.
This has been studied in detail in Ref.~\cite{Pecovnik_JNM2021} for the 0.23 dpa data point shown in Figure~\ref{fig:Concentration}. There the authors simulated the experimental results using a macroscopic rate equation code~\cite{Hodille_PhysScr2016}, using three defect types~\cite{Pecovnik_NF2020} with trapping energies corresponding to monovacancies and vacancy clusters.
As this modelling gives a good fit to the experimental data, without recourse to hydrogen trapping on other defects, we will also assume here that it is the vacancies which are most significant in relation to hydrogen retention.

Our approach will differ by how we determine the count of trapped deuterium (D) atoms per vacancy.
Previously, a great deal of work has gone into evaluating trapping energies for individual traps~\cite{Zibrov_2019}, and this information is invaluable for modelling outgassing as a function of temperature, such as in thermal desorption spectroscopy. Here our goal is only to determine the maximum retained deuterium. We instead use a model inspired by Hayward and Fu~\cite{Hayward_PRB2013,Hayward_JNM2016}. They showed, using density functional calculations, that hydrogen saturates the surface of vacancy clusters in $\alpha$-Fe before forming H$_2$ gas bubbles within, so that the important parameter was the void surface area. In tungsten this is likely also to be so, as in our simulations the vacancy clusters do not grow significantly. Given that the monovacancy is generally considered to trap up to 5 hydrogen atoms at room temperature~\cite{Pecovnik_NF2020}, and that these occupy 5/6 of the $[^1\!/_2 00]$ interstitial positions surrounding the vacant site, we can use the simple model that the surface area of a monovacancy is 5/6 occupied. For a general vacancy cluster with surface area $\Sigma$, computed using the isosurface method above, we say the D retention is
    \begin{equation}
        n_D = \frac{5}{6} \frac{\Sigma}{\Sigma_V},
    \end{equation}
where $\Sigma_V$ is the surface area of a monovacancy.

Figure \ref{fig:Concentration} shows the final result of our study, an estimation of the maximum deuterium concentration possible in our irradiated simulation cells, compared with direct NRA measurements of the concentration of D in irradiated tungsten. We see that the CRA only simulations greatly overestimate the saturation level, suggesting a D concentration over 10 at.\%.
This is understood as a consequence of the overestimation of vacancy-type defects in CRA due to the lack of relaxation processes.
The CRA+MD simulations however make an excellent estimation of the D retention, at order 1.5-2.0~at.\% in the saturation limit.

The most significant difference between the experimentally measured D concentration and our simulated estimate is the offsetting of the damage scale ($x$-axis). 
This is not unexpected in this case, as the simulations use a canonical measure of the number of vacancies inserted into the simulation, whereas the experiment uses a measure defined by counting the number of Frenkel pairs created in binary collisions assuming an uncertain threshold energy for this process. The experiment also can allow some additional long-time thermal relaxation of defects at room temperature which can not presently be simulated.

\section{Conclusions}

In this paper we have demonstrated how to generate converged highly irradiated microstructures to high doses with a combination of (static) CRA and molecular dynamics cascades.
In matching the definition of canonical displacements per atom we also match almost perfectly the defect content of MD cascade simulations and CRA+MD simulations, which in turn means we have proved the two simulation techniques are compatible.

We have shown that the Wigner-Seitz methodology for finding defects can be employed in these highly damaged cells, provided we take care.
We discussed how to generate the correct reference lattice, even when the simulation cell is sheared or rotated, using a very simple matrix inversion. We refined this reference lattice by taking into account the measurable homogeneous deformation, and adjusting the offset of the lattice motif.
We further showed that this standard measure of the occupation of lattice sites gives a good estimator of the point defect numbers, but if we want to discriminate between vacancies appearing as voids and those appearing in vacancy loops we can do this by finding isosurfaces in the Wigner-Seitz occupation.
By this means we were able to compute not just the number of vacant reference lattice sites, but rather the surface area of voids in the simulation.

In our simulations the relaxation is performed by MD cascades rather than "true" thermal annealing, and so there is little vacancy mobility, and little void growth. This is probably a reasonable assumption to compare to experiments performed at temperatures below the onset of vacancy mobility, as we have done here, but our methodology for generating microstructures will need to be supplemented by thermal annealing to compare to higher temperature experimental results.
There are few divacancies, and the surface area of a divacancy is just slightly under the surface area of two monovacancies.
Therefore the fractional surface area measured is, in this case, a close match to the fraction of vacant lattice sites.
Importantly we have shown that our methodology continues to work accurately even when voids do develop.

We have found that the surface area develops linearly with dose for low dose, but then flattens and saturates between 0.01 and 0.1 dpa.
Using a simple model for hydrogen retention, based on 5/6 of the possible vacancy cluster surface area being occupied, we get a very good match to the concentration of retained hydrogen implanted from a plasma into irradiated tungsten.

As our model for hydrogen retention is essentially parameter-free, depending solely on a robust estimation of the total vacancy surface area, we would expect it to reproduce the changes in hydrogen retention due to changes in microstructure observed by changing the elastic boundary conditions, or by introducing impurity atom types.

\section*{Acknowledgements} 
 This work has been carried out within the framework of the EUROfusion Consortium and has received funding from the Euratom research and training programme 2014-2018 and 2019-2020 under grant agreement No 633053 and from the RCUK [grant number EP/T012250/1]. To obtain further information on the data and models underlying this paper please contact PublicationsManager@ukaea.uk. The views and opinions expressed herein do not necessarily reflect those of the European Commission.
 We would like to thank P.-W. Ma for stimulating discussions.
Computer time granted by the IT Center for Science -- CSC -- Finland is gratefully acknowledged.

\bibliographystyle{ieeetr}
\bibliography{main}

\section{Appendix}

\subsection{Computing Local Roto-Strain field}
\label{sec:rotoStrain}

In this section we will describe the algorithm used to compute the local rotation and strain fields.
This is done with an iterative real space method.
If we know the lattice type, then from atom $i$ at position $\vec{x}_i$, we would expect to find its $k^{th}$ neighbour at position $\left(\vec{x}_i + \vec{v}^0_{k,i}\right)_{\mathrm{min}}$ if the lattice were locally perfect~\footnote{The subscript $_{\mathrm{min}}$ reminds us to use the minimum image convention when adding or subtracting vectors in a periodic supercell. Henceforth we will drop this subscript for clarity of notation.}.
In the perfect crystal, each atom in the same sublattice is expected to have neighbours at the same vector positions.
If we actually find that atom $i$ has a neighbour at a position $\vec{x}_i + \vec{v}_{k,i}$, where $\vec{v}_{k,i}$ is not the expected ideal lattice vector, then we can use this information to find the local strain. 
This is a simple task where strains are small, but around a dislocation core this may not be true and care must be taken.

The basis of the strain-finding algorithm is to minimise a fitting function of the form
    \begin{equation}
        \label{eqn:fitLocalStrain}
        S_i = \sum_k \left( \mathbf{T} \vec{v}^0_{k,i} + \vec{\delta} - \vec{v}_{k,i} \right)^2,
    \end{equation}
with respect to the nine matrix elements of the matrix $\mathbf{T}$ and the three elements of a uniform vector offset $\vec{\delta}$.
Differentiating equation~\ref{eqn:fitLocalStrain} with respect to these twelve elements gives a set of simultaneous linear equations, which can be solved with the Lapack routine DSYSV.
In our code we use a simple link-cell list~\cite{AllenAndTildesley} to find the neighbours of atom $i$.
This returns an unsorted list of neighbour vectors $\vec{v}_{j,i}$.
We sort this list by pairing each member of  $\vec{v}_{j,i}$ with the expected vector $\vec{v}^0_{k,i}$ which has the smallest separation $| \vec{v}_{j,i} - \vec{v}^0_{k,i} |$, rejecting any vector for which $| \vec{v}_{j,i} - \vec{v}^0_{k,i} |>a_0/4$ for all $k$, where $a_0$ is the lattice parameter.
With the list sorted we can minimise the fit function, but we could still find unexpected results if we applied equation \ref{eqn:fitLocalStrain} unthinkingly atom-by-atom, as our matching is making an implicit assumption that the strain is small.

To ensure a reasonably smoothly varying local strain, we note that we can add weighting and an initial guess to equation~ \ref{eqn:fitLocalStrain}. 
Let us imagine we have minimised $\sum_i S_i$ and found a global homogeneous strain $\mathbf{T}^0$ and offset $\vec{\delta}^0$ which best matches all the atoms simultansously.
We can now write $\vec{v}^1_{k,i} = \mathbf{T}^0\vec{v}^0_{k,i} + \vec{\delta}^0$, and pair the observed set of neighbours to this new set of expected positions. Then we can optimise
    \begin{equation}
        \label{eqn:weightLocalStrain}
        S = \sum_i w_i \sum_k\left( \mathbf{T} \vec{v}^1_{k,i} + \vec{\delta} - \vec{v}_{k,i} \right)^2,
    \end{equation}
where now $\vec{v}_{k,i}$ is a neighbour to atom $i$ paired with expected neighbour vector $\vec{v}^1_{k,i}$, and $w_i$ is a weighting for atom $i$.
This is the same set of simultaneous equations, but now each expected neighbour position is strained, rotated and shifted.

We can make a locally varying estimate of the strain by setting the weights $w_i$ with a Gaussian function centred on a point $\vec{x}$ and width $\sigma$.
    \begin{equation}
        w_i = \mathrm{Exp}\left[ -\frac{\left| \vec{x}_i - \vec{x} \right|^2 }{2 \sigma^2} \right].
    \end{equation}
Now minimising equation \ref{eqn:weightLocalStrain} finds the best strain locally to $\vec{x}$.
If $\sigma=\infty$, then each atom is weighted equally, and equation~\ref{eqn:weightLocalStrain} 
returns a slightly improved solution to the homogeneous strain given by the matrix product $\mathbf{T} \mathbf{T}^0$, and offset $\mathbf{T} \vec{\delta}^0 + \vec{\delta}$.
If our initial guess was good, we expect $\mathbf{T}$ to be close to the identity and $|\vec{\delta}|$ small.
 
To refine the local strains, we compute equation \ref{eqn:weightLocalStrain} on an evenly-spaced mesh of nodes spanning the supercell with spacing $\sigma$ set to the shortest supercell dimension, and $\vec{x}$ placed on each node in turn, taking the homogeneous solution $\vec{v}^1_{k,i}$ as our initial guess. 
This gives a new set of spatially varying strains $\mathbf{T}^1(\vec{x})=\mathbf{T}(\vec{x}) \mathbf{T}^0$, and displacements $\vec{\delta}^1(\vec{x}) = \mathbf{T}(\vec{x}) \vec{\delta}^0 + \vec{\delta}(\vec{x})$ on the nodes. Importantly, because we have chosen $\sigma$ to be large, this makes $\mathbf{T}(\vec{x})$ near unity and slowly varying.

We then make a linear interpolation of strains and displacements to a new mesh of nodes with spacing $\sigma/4$, to seed a new spatially refined estimate of the expected atom positions, $\vec{v}^2_{k,i}(\vec{x}) = \mathbf{T}^1(\vec{x}) \vec{v}^1_{k,i} + \vec{\delta}^1(\vec{x})$ .
This process is iterated, each time reducing the mesh spacing, computing a small local change in the strain and displacement, and the small local change in the expected positions of the neighbours.
By this process we can build up a potentially large local transformation from small incremental steps.
We stop the iteration at the $n^{th}$ level when $\sigma \approx a_0$, and we have an evenly spaced mesh of nodes with the spacing of the lattice parameter.

Note that the derivative of the local displacement vector $\vec{\delta}^n$ produced at the end of the iteration is not related to the local strain- these displacements are only used to match observed neighbours to expected neighbours and improve the fit for the strain. We discard it.

The final local rotostrain transformation $\mathbf{T}( \vec{x} )$  at a general point $\vec{x}$ used in equation~\ref{eqn:rotostrain_normal} is needed as a continuous field, so is taken to be the linear interpolation of the final iteration $\mathbf{T}^n(\vec{x})$ computed on the eight nodes nearest to $\vec{x}$.

\end{document}